# Blueprinting quantum computing systems

Simon J. Devitt

Centre for Quantum Software and Information, University of Technology Sydney, Australia

Email: simon.devitt@uts.edu.au

## Abstract

The development of quantum computing systems has been a staple of academic research since the mid-1990s when the first proposal for physical platforms were proposed using Nuclear Magnetic Resonance and Ion-Trap hardware. These first proposals were very basic, essentially consisting of identifying a physical qubit (two-level quantum system) that could be isolated and controlled to achieve universal quantum computation. Over the past thirty years, the nature of quantum architecture design has changed significantly and the scale of investment, groups and companies involved in building quantum computers has increased exponentially. Architectural design for quantum computers examines systems at scale: fully error-corrected machines, potentially consisting of millions if not billions of physical qubits. These designs increasingly act as blueprints for academic groups and companies and are becoming increasingly more detailed, taking into account both the nature and operation of the physical qubits themselves and also peripheral environmental and control infrastructure that is required for each physical system. In this paper, several architectural structures that I have worked on will be reviewed, each of which has been adopted by either a national quantum computing program or a quantum startup. This paper was written in the context of an award with the Royal Society of New South Wales, focused on my personal contributions and impact to quantum computing development, and should be read with that in mind.[1]

## Introduction

The development of the second generation of quantum technology such as quantum computing platforms, quantum communication systems, quantum simulators and sensors is anticipated to create a new technological revolution, similar to the digital computing revolution of the 20th century. While significant research and development in quantum technology began in the middle of the 1990s, largely in the academic space, since 2014 we have seen an explosion in investment and consequently an explosion of progress.

Governments, multi-national corporations and the equity investment community have recognised the potential impact of quantum technology and have invested accordingly. A €1 billion European flagship (Riedel et al. 2019), the US$1.2 billion quantum initiative (Raymer and Monroe 2019), over US$10 billion in declared investment from the Chinese government (Zhang et al. 2019), a €2 billion quantum investment from the Germans as part of their COVID-19 recovery package, and others (Brennan et al. 2021), have cemented quantum technology as a major global development goal in the 21st century. Corporate investment in building, particularly, quantum computing systems is also extensive: Google, Microsoft, Amazon, IBM, Baidu, TenCent and Alibaba are just some of the major global technology firms that have established extremely strong research groups to build and deploy quantum computing systems. IBM and Amazon have already launched fee-based cloud access to several quantum computing platforms, such as Ion-Traps and Superconducting systems. Quantum startups are now ubiquitous worldwide, both in the hardware and software space, with the three largest quantum computing startups — IonQ[2], Rigetti[3] and PsiQuantum already valued over one billion dollars.[4]

---

[1] In 2021, Simon Devitt was awarded the inaugural Warren Prize for research by engineers and technologists in their early to mid-careers, by the Royal Society of New South Wales.

The construction of a large-scale quantum computer is now a serious goal amongst national programs, multinational corporations and the equity funding community and there is a variety of physical platforms that people are developing. Which platform or platforms will end up being viable for scientifically or commercially useful computational tasks is still up for debate, but the diversity of approaches will undoubtably spur more rapid development.

Over the past 15 years, I have been involved in several results related to quantum architecture development and design, with numerous exceptional theoretical and experimental colleagues. We introduced the one of the earliest large-scale quantum computing architectures that incorporated quantum error correction, was modular in nature, and could be conceptually scaled to an arbitrary degree (Devitt et al. 2009) and have worked on architecture designs for multiple different physical systems (Oi, Devitt, and Hollenberg 2006; Stephens et al. 2008; Nemoto et al. 2014; Lekitsch et al. 2017; Mukai et al. 2020). We introduced a design for a high-performance quantum computing system that could perform distributed blind quantum computing in an error-corrected environment (Devitt, Munro, and Nemoto 2011) and designed architectures for quantum communication networks that could serve to connect distributed quantum computing systems together (Munro et al. 2010, 2012; Devitt et al. 2016).

Aside from architecture development, I have also developed parts of theoretical frameworks for how to program, implement and resource-optimise fully error-corrected quantum algorithms (Devitt, Munro, and Nemoto 2013). This includes examining the practical requirements of classically processing error-correction information from a quantum computer (Devitt et al. 2010), how to compile high-level quantum circuits into error-corrected compatible forms (Fowler and Devitt 2012; Herr et al. 2018; Herr, Nori, and Devitt 2017; Paler et al. 2014; Devitt 2016; Horsman et al. 2012), and what the formal requirements are for benchmarking quantum algorithms on practical machines (Devitt et al. 2013; Meter and Devitt 2016; Paler, Herr, and Devitt 2019).

In this paper, I specifically examine three of the scalable quantum computing blueprints that I have been involved in that have been adopted by quantum startup companies and national programs worldwide. I look specifically at three designs: one in Ion-Traps that has been adopted by the UK startup Universal Quantum (Lekitsch et al.

---

2  https://finance.yahoo.com/quote/IONQ/

3  https://www.rigetti.com/merger-announcement

4  https://www.wsj.com/articles/psiquantum-raises-450-million-to-build-its-quantum-computer-11627387321

2017), one using Nitrogen Vacancies (NV) in diamond, adopted by the US-based startup Turing inc and the Austrian start-up Godel GmbH (Nemoto et al. 2014), and one in superconductors that has been adopted by the Japanese national program, Q-Leap and Moonshot (Mukai et al. 2020; Kwon et al. 2020).

## The basics of quantum computing

The core operational element of a quantum computer is the quantum bit (qubit): this is a well-defined two-level quantum system that can exist in a variety of physical platforms. One of these two levels corresponds to a binary 0 state and the other to a binary 1 state (Nielsen and Chuang 2000).

The physical systems used to define these two states can be anything from the electronic levels of an ionised atom (Cirac and Zoller 1995), the polarisation state of a single photon (O'Brien 2007) or the spin state of a phosphorus atom in a silicon crystal (Kane 1998)[5].

A qubit lives in a two-dimensional complex vector space, where the general state of a qubit is given by equation 1:

$$|\psi\rangle = \alpha|0\rangle + \beta|1\rangle, \quad (1)$$
$$|\alpha|^2 + |\beta|^2 = 1, \quad \{\alpha, \beta\} \in \mathcal{C}$$

Gate operations on qubits are defined by unitary operations on this vector space. As the modulus squared of all amplitudes sums to one — as they represent the probabilities that qubits are measured in one of the two possible basis states — unitarity is required to ensure that probabilities are conserved as quantum gates are performed.

When considering an array of qubits in a quantum computer, the total size of the vector space grows exponentially. For an $N$-qubit quantum computer, the complete state of the system can be described by a complex column vector of size, $2N$:

$$|\psi\rangle = \begin{pmatrix} z_0 \\ z_1 \\ z_2 \\ . \\ . \\ . \\ z_{2^N-1} \end{pmatrix}, \quad \sum_{i=0}^{2^N-1} |z_i|^2 = 1. \quad (2)$$

Gates are then defined as unitary matrix operations, $G$, of size $2N \times 2N$, where $G^\dagger G = I$, and $\dagger$ is the conjugate transpose of $G$.

This is effectively what a quantum computer is. It is a matrix multiplier over a complex column vector of size $2N$. The inability for classical computers to simulate a quantum computer is due to the exponential scaling of this column vector. As each element in the matrix is a complex number, and each complex number requires two real numbers (doubles), the memory required to completely store the state of a quantum computer containing $N$ qubits is (in Bytes) $2 \times 8 \times 2^N$.

Even for small qubit arrays, this scaling overtakes the memory capacity of any classical system. For example, the Google Sycamore chip contains $N = 53$ functional qubits (Arute et al. 2019). To completely store the state of Google's Sycamore chip would require 144 PetaBytes. This is essentially the argument being used in a recent IBM paper claiming a method to simulate Google's quantum supremacy has resulted in the Summit supercomputer (Pednault et al. 2019). The argument is to re-task hard-disk space as virtual RAM to reach the capacity needed to store the entire complex vector representing the Sycamore quantum processor.

---

5 This paper inspired Michelle Simmons FRS FRSN. [Ed.]

There are multiple techniques that can be used to approximate the behaviour of a quantum computer in classical systems and there are efficient classical algorithms for exact simulation of restricted classes of gate operations (Aaronson and Gottesman 2004; Markov and Shi 2008; Perez-Garcia et al. 2007). However, we know that the efficient simulation of a universal set of quantum gates is not possible unless fundamental conjectures of complexity theory are proven to be false (Aaronson 2013).

Qubits naturally existing in this complex vector space implies that a quantum computer only requires $N$ qubits to perform the same computation, and a variety of results since the early 1990s have demonstrated that the complex vector space of quantum logic is more powerful than binary logic for a specific set of problems. This includes direct simulation of quantum systems (Lloyd 1996), solving the hidden abelian subgroup problem (Kitaev 1995) — which includes factoring large integers (Shor 1994) — solving numerous optimisation and graph problems (Lee, Santha, and Zhang 2020), and solving large sets of linear equations (Harrow, Hassidim, and Lloyd 2009).

Quantum computing is now growing into a larger and larger industry, with numerous multinational technology companies constructing and making available small quantum processors over the cloud. National initiatives being established in most major industrial countries (Roberson and White 2019; Riedel et al. 2019; Sussman et al. 2019; Zhang et al. 2019; Yamamoto, Sasaki, and Takesue 2019; Fedorov et al. 2019; Raymer and Monroe 2019; Knight and Walmsley 2019) and investment from venture capital firms into both quantum hardware and software companies now eclipsing the $1 billion scale.

Part of the development of quantum computers is asking how we blueprint large-scale quantum computing systems — to the level needed to implement the wide variety of quantum algorithms that are provably more efficient than their classical counterparts. These blueprints include the hardware design of the qubits and qubit control systems, the manner in which Quantum Error Correction (QEC) is embedded within these designs, how environmental infrastructure is deployed for these systems, and how the classical computing support structure is integrated to the quantum hardware.

## The core elements of an architecture: the DiVincenzo criteria

David DiVincenzo was one of the first to enumerate a minimal list of physical requirements that are needed to build a large-scale quantum computer. In 2000 he published what are now referred to as the DiVincenzo criteria (DiVincenzo 2000): a set of five elements a quantum system must have to be, in principle, suitable for constructing a quantum computer. These are:

1. A scalable physical system with well-characterised qubits
2. The ability to initialise the state of the qubits to a simple fiducial state
3. Long relevant decoherence times
4. A "universal" set of quantum gates
5. A qubit-specific measurement capability.

These five conditions are a necessary minimal set that is required for a physical platform to be appropriate for a quantum com-puter, but they are by no means sufficient when building a scalable system.

From the middle of the 1990s to approximately the middle of the 2000s, there were literally dozens of physical quantum systems proposed that satisfied the DiVincenzo criteria. Many of these "architecture" proposals have slowly disappeared as researchers have realised that the details of a practical quantum architecture go well beyond these five elements.

Of the many different physical systems proposed for quantum computing, there remain eight that are under major development:

1. Superconducting qubits (Kwon et al. 2020)
2. Ion Traps (Brown, Kim, and Monroe 2016; Lekitsch et al. 2017)
3. Optical qubits (single photon and continuous variables) (O'Brien 2007; Braunstein and Loock 2005)
4. Colour centres (such as Nitrogen Vacancy centres in diamond) (Nemoto et al. 2014)
5. Quantum Dots (Jones et al. 2012)
6. Donors in Silicon (Kane 1998)
7. Neutral Atoms (Saffman 2018)
8. Anyonic Systems (Nayak et al. 2008).

A set of criteria that I generally use to define a "major" quantum computing system are:

1. There is significant funding available for the platform.
2. The platform has already demonstrated the fabrication and control of a small number of qubits (1–10) to the point where it is now somewhat routine.
3. There are experimental and/or theoretical researchers involved in a systems development that are "true believers," i.e. they are focused strongly on actually building a scalable quantum computer, rather than simply doing interesting and more foundational physics work.

Each of the eight systems listed above satisfy each of these criteria, except arguably for Anyonic systems, but the vast amount of money invested by Microsoft into this highly experimental platform necessitates its inclusion on the above list. Investment into these platforms is somewhat evenly distributed across corporations, governments, universities and startups.

## Modern quantum architecture designs

Beyond the DiVincenzo criteria, architecture development and blueprinting quantum computing systems have evolved significantly over the past 10–15 years. As we are entering the era of engineering small qubit chipsets, designs for larger-scale systems are becoming more complex. Early quantum computing blueprints generally consisted of little more than identifying an appropriate two-level quantum system and describing the interaction dynamics that enabled a universal set of quantum gates, initialisation and measurement.

As we have further understood the actual necessities of a large-scale quantum computer, design blueprints across a variety of different systems have become more detailed and more sophisticated. The most notable change is the detailed introduction of Quantum Error Correction (QEC) protocols.

## Quantum error correction

Fabricating and controlling physical qubits is a difficult thing to do. The coupling of individual qubits to the environment inevitably leads to loss of quantum coherence (known as decoherence). From an information-processing standpoint, this decoher-ence effect introduces errors, which quickly renders the output from any quantum algorithm essentially random. Errors can come from inaccurate fabrication, bad qubit design — where the environment couples too strongly to the two-level system that defines the qubit — or could be induced by imperfect control.

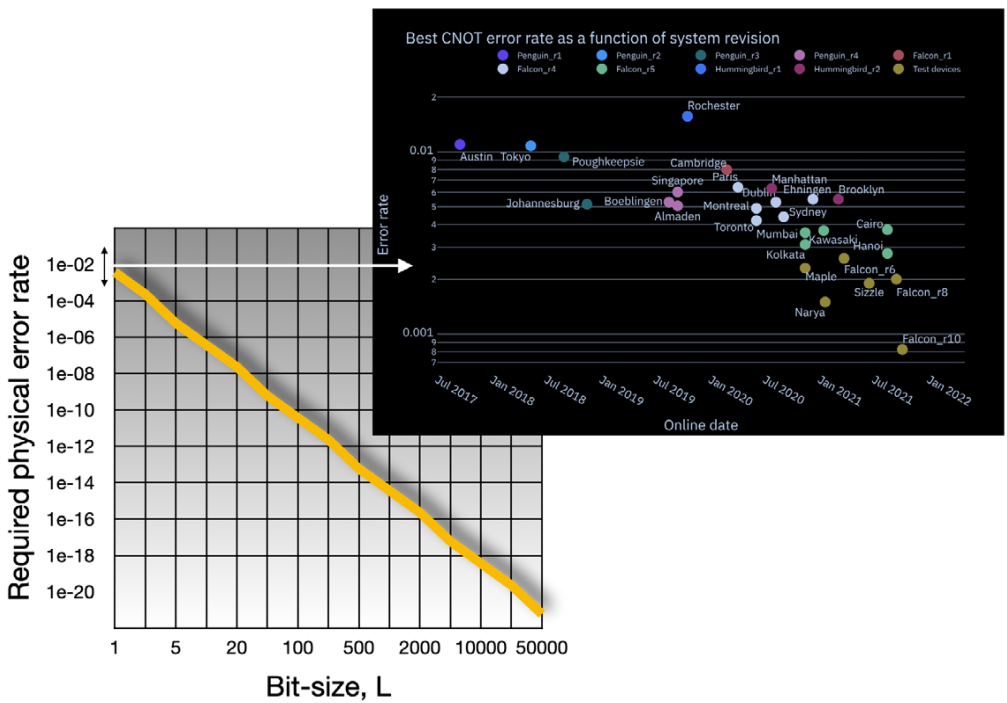

Figure 1: Plot of the required error physical qubit error rates needed to implement Shor's algorithm for various bit sizes, *L*, using the construction of Gidney and Ekera (2019). The insert illustrates experimental error rates over IBM's deployed superconducting systems between 2017 and 2021 (superconducting qubits were first demonstrated in the late 1990s, with error rates on the order of 5–10%). Experimental error rates only live in the upper left-hand corner of required error rates for Shor's algorithm.

While experimentalists have done an impressive job at decreasing qubit error rates in physical systems over the past two decades, physical engineering alone will not be sufficient to reduce error rates to a degree necessary for large-scale quantum algorithms to be run. Shown in Figure 1 is a plot of the physical error rates needed for qubits if you were to implement Shor's factoring algorithm.

For Shor's algorithm, we wish to be able to factor numbers that have a bit-size length, *L*, of *L* > 1024, as this is approximately the minimum key size used in modern implementations of RSA public key encryption. From Figure 1, you can see that this would require physical error rates on the qubits of at most $O(10^{-14})$.

Shown in the insert of Figure 1 is a plot of the physical error rates achieved in the laboratory since qubits were first fabricated and tested in the late 1990s. While significant progress has been made, the error rates achieved in the lab insert lie in the very top left-hand corner of the larger plot in terms of the physical error rates that have been demonstrated, and that experimentalists have reduced these errors from approximately 3–5% to approximately 0.1% in 20 years. This does not prove that a new, revolutionary method for qubit fabrication and control will not be made that allows physical errors to drop by a further 10 or 11 orders of magnitude, but it does suggest that physical systems are going to need some help when it comes to reducing error rates to the level that is needed to implement large-scale quantum computing.

Quantum Error Correction (Devitt, Munro, and Nemoto 2013) provides this framework, by taking high error rate, physical qubits and encoding them into a code block to form a logical qubit. This logical qubit then contains sufficient redundancy so that errors on the physical qubits can be detected and corrected without destroying or unintentionally modifying any information within the encoded block.

While there are a plethora of QEC codes to choose from, in terms of large-scale quantum architecture design there has been unarguably a preferred technique, known as the surface code (Fowler et al. 2012). The surface code encodes a single logical qubit into a two-dimensional array of physical qubits. Illustrated in Figure 2 is an encoded qubit using a distance $d = 5$ code.

Code distance is a measure that counts the minimal number of physical errors that is required to create a logical error. i.e. if we consider a logically encoded $|0\rangle_L$ state, how many physical bit-flips ($X$-gates) are required in order to take $|0\rangle_L \leftrightarrow |1\rangle_L$? For a distance $d = 5$ code, we require five physical $X$-gates to induce a logical bit-flip.

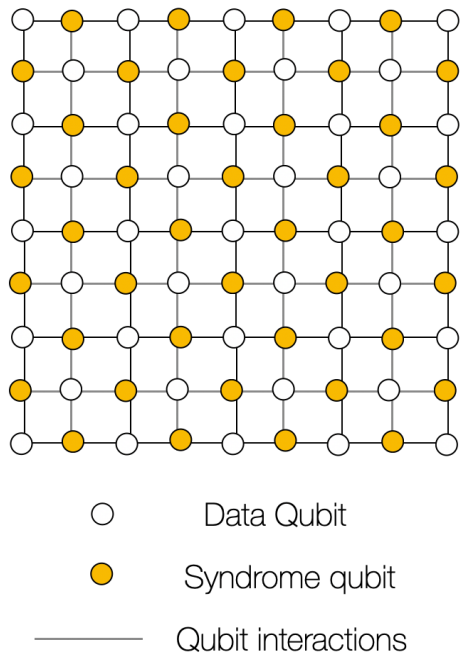

Figure 2: A 2D lattice of qubits encoding a distance surface code logical qubit. A distance code requires a lattice of $(2d - 1)^2 = 9 \times 9 = 81$ qubits and is able to correct for up to two arbitrary errors on any one of the physical qubits. Approximately half of the qubits are actually part of the encoded block (coloured white) and the others are Syndrome qubits (yellow) which are used to extract error information and are repeatedly measured. The solid likes represent the required interactions between qubits. All qubits must be interacted with their four nearest neighbours.

In quantum information we need to protect against both bit-flip gates ($X$-gates) and phase-flip gates ($Z$-gates) which can cause errors in quantum superposition, i.e.:

$$(|0\rangle + |1\rangle)/\sqrt{2} \quad \leftrightarrow \quad (|0\rangle - |1\rangle)/\sqrt{2}.$$

In the most general case, it is assumed that the probabilities of physical bit-flips and phase flips are identical and hence we need identical levels of protection for these two types of errors in the logical qubit. In the surface code, this is achieved by ensuring that the surface code is a square lattice.

We will not summarise the specific details of how error-correction works, there are sufficient reviews in the past 20 years to cover these topics (Devitt, Munro, and Nemoto 2013; Fowler et al. 2012). However, we will briefly discuss how the encoded error rates (logical error rates) scale.

Through direct numerical simulation of the error correction protocols on the surface code (Devitt et al. 2016), it is possible to relate the distance of the code, $d$, the physical error rate associated with each qubit in the code, p, and the failure probability of the logically encoded information, $P_L$, as equation 3:

$$P_L \approx C_1(C_2 p)^{\frac{d+1}{2}}. \quad (3)$$

These simulations track the effect of Pauli bit and phase errors on the quantum circuits used to implement the circuit code, apply error-decoding protocols and then estimate whether logical Pauli errors are formed. By Monte Carlo simulations of many instances, it is possible to estimate the logical failure rate of the encoded qubit, $P_L$.

The first thing to notice is that as $d \to \infty$, the logical error rate $P_L \to 0$ if and only if $C_2 p < 1$. If $C_2 p > 1$, the logical error rate gets worse as a function of code distance. This is what is known as the code threshold, i.e. the largest physical error rate that the code can still provide correction. If physical error rates are larger than the threshold, errors are introduced faster than they can be extracted by the error correction and logical error rates get worse instead of better.

The fault-tolerant threshold for an actual architectural model is heavily dependent on the type of error-correction code used, the type of errors that are anticipated at the physical later in a specific hardware system, and how a hardware model physically realises the gate operations needed to realise the chosen code. The actual threshold for the surface code is more explicitly modelled in numerous papers and has been found to be approximately (Fowler et al. 2012) under a balanced Pauli noise model, assuming an underlying architecture based on a 2D nearest-neighbour array of interacting qubits.

For the surface code, the code distance is related to the number of physical qubits in the lattice, $N$. Specifically a code of distance d requires a $N = (2d-1)^2$ qubit square lattice. Hence we can rewrite eq. 3 as:

$$P_L \approx C_1(C_2 p)^{-\frac{\sqrt{N}+3}{4}}. \quad (4)$$

Consequently, if $p$ is under the code threshold, the logical error rate $P_L$ will decrease, exponentially, with the $\sqrt{N}$, which is the number of physical qubits along the edge of the 2D lattice that defines a surface code logical qubit.

This is the mechanism that allow us to reduce the logical error rate of an error-corrected qubit without having to find ways to reduce the physical error rates of constituent physical qubits. Instead of performing Shor's algorithm on physical qubits which have been engineered to have error rates on the order of $p \approx 10^{-14}$ to factor numbers larger than 1024 bits, we instead create sufficiently large encoded qubits such that Eq. 4 achieves a logical error rate that is small enough. This does not require us to change $p$ (provided it is already below the threshold), but rather it requires us increase $N$, i.e. build more physical qubits.

Hence, large-scale quantum computation is largely an exercise in building an error-correction machine. As one of the pioneers of error-correction, Andrew Steane, once reportedly quipped:

> A quantum computer is an error-correction machine, computation is just a by-product.

This is reflected in large-scale design blueprints. The majority of the design and analysis is focused around implementing error-correction for a large number of qubits in the most effective manner possible.

It should be heavily stressed that error-correction adds a very large overhead for any quantum algorithm. This is not only due to the direct overhead of encoding logical qubits with a collection of physical qubits, but is also due to ancillary protocols needed to maintain fault-tolerance for encoded operations.

This consequently means that many algorithms that provide a clear quantum advantage over classical computers require millions if not hundreds of millions of qubits. Most of the most accurate benchmarking of quantum algorithms have occurred for factoring (Gidney and Ekera 2019; Devitt et al. 2013), quantum simulation (Reiher et al. 2017; Babbush et al. 2018) or applications using quantum random-access memories (Matteo, Gheorghiu, and Mosca 2020).

## Infrastructure and control

Another major component that has become a staple of blueprinting quantum computing systems is related to the classical control systems and environmental infrastructure needed to operate specific qubit systems.

As quantum information is so fragile, and environmental decoherence can couple into a system so easily, it is often required that physical qubits need to be housed in infrastructure that carefully controls the environment in which the qubit lives. This can take multiple forms, depending on the specific system being designed.

In some systems, qubits need to be kept at incredibly low-temperature environments. This is particularly true for superconducting qubits or qubits built from phosphorus donors in silicon or qubits built from artificial atoms (quantum dots). These qubits are often characterised by energy separations that occur at microwave frequencies. At these frequencies, thermal photons from the environment can be of the same order as the energy of the qubit. Consequently they can induce qubit dynamics that are otherwise unwanted. i.e. noise and errors.

As many of these qubits are designed to operate at these energy separations, the only other choice is to make the environment as cold as possible as to suppress the production of these thermal photons, which is exactly what is done.

Dilution refrigeration systems are now commonplace around the world, which allows for the reduction of temperature in a sample chamber to the milli-Kelvin range. The issue with dilution refrigeration cooling is that it significantly constrains the number of qubits that can physically be placed inside such a refrigerator. It also constrains the amount of energy that can be routed into a refrigerator to perform qubit control, initialisation and measurement without overwhelming the ability of the refrigerator to keep the system cold. This requires significant thought as to how to overcome what are essentially engineering roadblocks to qubit systems that may contain millions or billions of physical qubits.

In other systems, such as Ion-Traps, qubits need to be placed into an essentially perfect vacuum. An ion-trap is an electromagnetic trap that can be used to confine a single charged atom. Clearly, such a trap becomes ineffective in an atmosphere, where collisions between atmospheric atoms and the trapped ion can easily eject the ion from the electromagnetic trapping potential. Consequently, Ion-Trap quantum computers are routinely pumped down to a pressure of approximately $10^{-11}$ Torr or better. This requires an initial pumping down of the chamber, a cooking stage above 200K to cause additional atoms stuck within the walls and other surfaces to evaporate into the chamber, and then a second pumping stage. Additionally, it is expected that for a scalable system, the vacuum system will need to be cooled to 4K to further reduce the rate at which ions are heated (Niedermayr 2015).

As with superconductors, environmental infrastructure can make scalability difficult. Having large vacuum systems that need to house millions or hundreds of millions of physical ions for long-term operation presents a difficult engineering challenge. Although, unlike superconductors, Ion-Traps have the ability to be connected with photons. This allows fibre optic coupling of independent ion-traps (Monroe et al. 2012), something that is extremely difficult to do in superconducting systems (Magnard et al. 2020).

Photonic computers — where qubits are individual photons — also have their own specific challenges. As photons travel at the speed of light, it is extremely hard to build computational chips, as you cannot spatially localise photons to do actual gate operations. While there are techniques that are used to overcome the problem that individual quantum bits are flying around your computer at 30 cm per nanosecond (Rudolph 2017), having physical qubits moving around this quickly creates significant challenges for scalability for these particular models.

Other physical platforms for quantum computing also have associated infrastructure and control issues, each of which has to be handled in its own way. Good quantum architecture designs and blueprints contain significant details with respect to achieving scalability in infrastructure and control, not just in terms of being able to fabricate a large number of individual qubits.

**Fabrication and cost**

The digital computing industry has unarguably been dominated by issues surrounding fabrication and cost. One of the most foundational principles in classical microprocessor development was Moore's law, an empirical prediction made in 1965 by the co-founder of Intel, Gordon Moore, who observed that the number of components on a classical microprocessor was doubling every 18 to 24 months.

This observation was not only related to the ability of fabricating more components onto an integrated circuit, but also the cost in doing so. Feature sizes and hence transistor sizes decreased exponentially at the same time as costs per transistor dropped exponentially. This is the primary reason behind the ubiquitous nature of information-processing technology worldwide.

While there have always been peripheral discussions with certain platforms about the ease and cost associated with the production of a quantum computer, these discussions have become much more serious as the technology moves out of the academic laboratories and into commercial production.

Certainly the most dominant issue that motivates discussions surrounding mass manufacturing and costs is current estimates surrounding the number of physical qubits needed to implement large-scale algorithms such as Shor's algorithm or problems in quantum chemistry. In Gidney and Ekera (2019) and Reiher et al. (2017) we see some of these estimates, which attempt to incorporate and optimise the full integration of error-correction and fault-tolerant protocols — given these resource estimates, the cost of a quantum computer to implement these algorithms can be estimated assuming various order-of-magnitude estimates for the Price Per Qubit (PPQ). These algorithms require a very large number of physical qubits, and hence cost becomes a big factor.

Table 1: Cost of various machines for well known quantum algorithms as a function of PPQ.

| PPQ | Factoring (Gidney and Ekera 2019) | Nitrogenase (Reiher et al. 2017) |
|---|---|---|
| $1000 | $20 Billion | $200 Billion |
| $1.00 | $20 Million | $200 Million |
| $0.01 | $200,000 | $1 Million |

Qubit costs can be parameterised as the PPQ, where you average over the total cost of the machine, including classical control systems, environmental infrastructure etc, as a function of the number of qubits in this system. For example, a superconducting system housing 50 actual qubits would require a $500K dilution refrigeration system and an additional $500K in microwave signal generators, niobium wiring, chip fabrication costs and classical computer control, and would have a PPQ of $20,000. At this scale, a quantum computer capable of factoring (Gidney and Ekera 2019) or the simulation of complex molecules such as nitrogenase (Reiher et al. 2017) would carry a price tag higher than the GDP of Austria ($417 billion) and Japan ($4.8 trillion). Table I illustrates the cost for factoring and quantum chemistry for various orders of magnitude for a PPQ. Even at an effective $1 PPQ, a quantum computer of sufficient size for factoring or quantum simulation would be a significant investment.

The entire quantum technology industry is founded on the precept that qubits will eventually be cheap and quantum technology will be at least as ubiquitous as large computational servers, if not as ubiquitous as mobile phones and the internet of things. To achieve this, PPQs must be reduced by at least five orders of magnitude compared to the state of the art today, and likely much much lower.

### Three hardware architectures

In this section I will summarise some of the architectures that I have been involved with (Nemoto et al. 2014; Lekitsch et al. 2017; Mukai et al. 2020; Kwon et al. 2020) and the development of the key theoretical results that allow for error-corrected implementations of quantum algorithms on these, and other, hardware (Devitt et al. 2010; Fowler and Devitt 2012; Fowler, Devitt, and Jones 2013; Herr, Nori, and Devitt 2017; Horsman et al. 2012; Paler et al. 2014, 2012; Paler, Devitt, and Fowler 2016; Devitt 2016; Herr et al. 2018; Paler and Devitt 2018).

#### Ion-traps

Many groups have proposed large-scale ion trap quantum computers (Cirac and Zoller 2000; Kielpinski, Monroe, and Wineland 2002; Schaetz et al. 2004; Duan et al. 2004; Metodi et al. 2005; Steane 2007; Stock and James 2009; Kim and Kim 2009; Crick et al.

2010; Amini et al. 2010; Monroe et al. 2012).[6] Two themes dominate — multi-zone micro-fabricated traps, with each zone containing a handful of ions, and optical linking of traps. Optical linking is probabilistic, with a success probability of 2.2 × 10$^{-4}$ the best achieved to date (Stephenson et al. 2020).

While this is expected to improve, the current state-of-the-art is far from sufficient for a practical computer (Nickerson, Li, and Benjamin 2013). Even if optical linking is improved by several orders of magnitude, this approach, while technically scalable, will result in a computer much slower than is theoretically possible. A direct interaction between ions separated by microns will generally be far faster than a multi-metre photon mediated interaction. Existing multi-zone micro-fabricated trap designs make use of control electrodes in the same plane as the trap electrodes, precluding scaling to an arbitrarily large 2-D lattice of qubits as the required number of control electrodes per unit length around the edge of the chip grows without bound. An alternative approach is required to achieve high performance and scalability.

Gven ions can be reliably transported along linear traps and through X-junctions (Walther et al. 2012; Wright et al. 2012), the core of our solution to the scaling problem is to abandon the optical linking of chips and instead align chips each containing a moderate number of X-junctions so that ions can be transported from chip to chip as though along an unbroken trap. Each X-junction under normal circumstances contains just two ions, only one of which is used as a qubit. The second ion would be used for sympathetic cooling (Rohde et al. 2001). A surface-electrode single X-junction is shown in Figure 3. The design incorporates electrodes to rotate the trap axes, enabling laser cooling of ions at any location (Stenholm 1986). This is necessary to enable characterisation of the junctions and adjustment of electrode control pulses at time of construction.

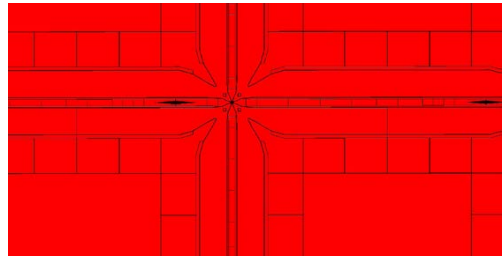

Figure 3: Schematic 5×5mm surface-electrode X-junction produced by the Sussex group in our initial design. Black diamonds represent holes above which ions are interacted and manipulation. Three optical fibres are cemented into each hole, two of which deliver laser light and the third of which is used to gather photons for measurement. A total of 52 DC connections are required per junction.

Each junction would occupy a 5 × 5 mm patch of a larger wafer and be associated with a single interaction and manipulation region. 16 of these junction patches would be bundled together to create a repeating 20 × 20 mm section (Figure 4). 25 of these sections would be fabricated on a 100 × 100 mm chip.

In addition to two Rf Voltage connections, each section also requires 48 fibre connections (3 connections per junction zone, two for the entanglement lasers one for detection) and 840 DC connections (52 connections per junction and 8 additional per section) for operation. An in-vacuum

---

[6] The work in designing an ion-trap architecture was performed with experimentalists from the University of Sussex, Google and researchers in Denmark and Germany (Lekitsch et al. 2017).

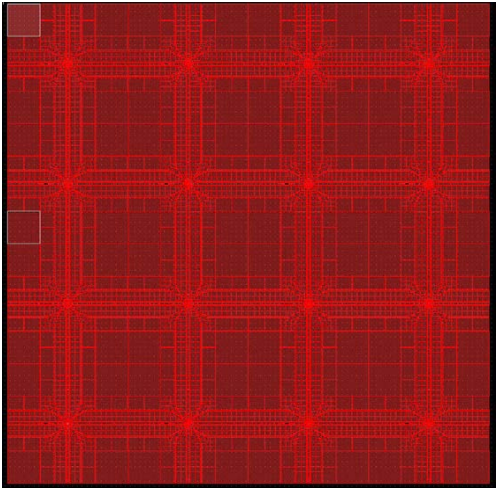

Figure 4: Schematic of a section of a larger wafer containing 16 X-junctions produced by the Sussex group in our initial design. This section would be tiled across the wafer. Each section would be electrically isolated from all others

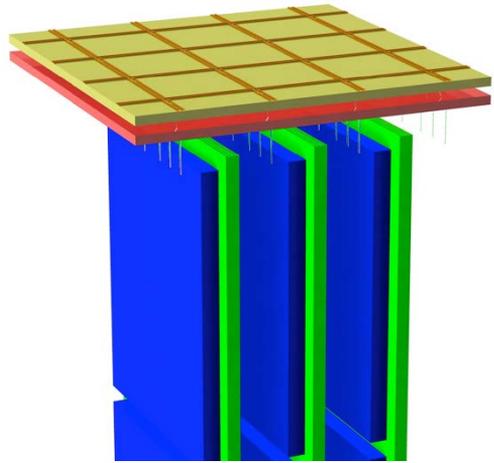

Figure 5: 20 × 20mm 16 X-junction section with chip carrier to gather the 840 DC electrode connections into neat rows of 100 × 100μm pads, 48 optical fibres, and three printed circuit boards schematically holding all required electronics (DACs, deserializers, amplifiers, filters, LC resonant circuit) within the footprint of the section.

DC system consisting of 21, 40-channel DACs (digital to analogue converters) is used to generate 840 different DC voltages, and reduces the number of DC connections required for each section to a total of 8. The 20 × 20 mm footprint is large enough to mount DACs, deserializers, amplifiers, filters, LC resonant circuit and fibres underneath the traps on vertical in-vacuum PCBs. This is shown schematically in Figure 5.

The individual 100 × 100 mm chips also need to be accurately aligned with one another to allow shuttling of ions from one chip to the other. If a minimum alignment precision of 5μm in each direction of the rails is achieved, fast adiabatic shuttling was shown to be possible in our simulations (Lekitsch et al. 2017). To achieve this alignment accuracy, a precision machined stainless steel frame would be mounted on top of a six-axis piezoelectric positioner, possibly with the frame topped with piezoelectric pistons to permit any warping of the chip to be corrected (Figure 6). The alignment could be performed during assembly of the chips, using a laser measurement system and microscope.

*Vacuum chamber*

The primary design constraints on the vacuum chamber are appropriate laser access and light removal, electromagnetic shielding, unobstructed and close line of sight to all traps for characterization, clean loading of ions so the trap is not degraded over time, and possibly low operating temperature.

Given light is delivered by fibres to every X-junction, it is not feasible to always have a direct line of sight through a window out of the vacuum chamber. Instead, we propose coating the inside of the vacuum chamber with a product such as Magic Black[7], which absorbs 99.99% of light at typical ion manipulation wavelengths. Excellent electromagnetic shielding and cryogenic operating temperatures have been achieved by using a thick-walled, oxygen-free, high-conductivity copper chamber in a helium-bath cryostat (Brown et al. 2011).

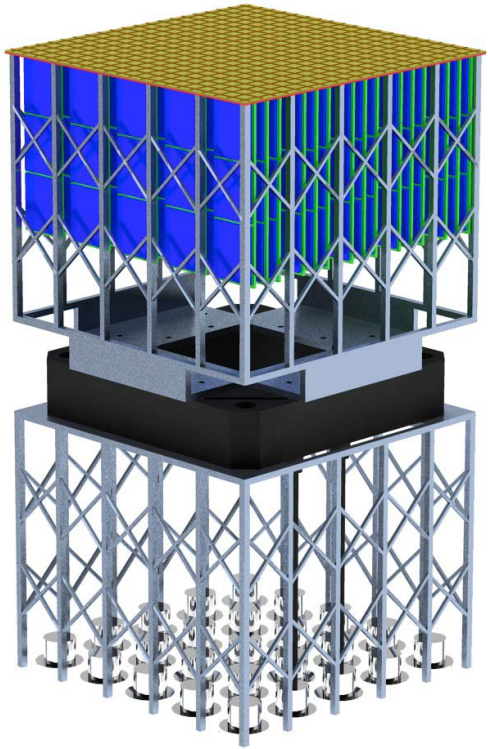

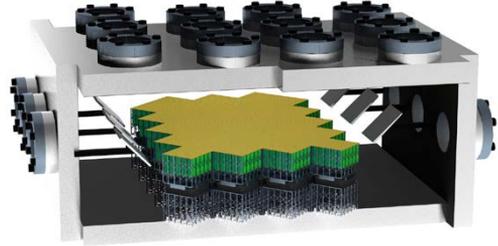

Figure 7: Repeating unit cell of a corridor of a high-performance surface code optimized ion-trap quantum computer. Windows in the side walls permit cooling laser sheets to enter. Windows in the roof next to the side walls permit exit of this light after reflection by 45° mirrors. Windows across the remainder of the roof permit characterization of the trap at time of construction.

Figure 6: 100 × 100 mm chip with underside mounted electronics, precision stainless steel mounting frame, six-axis piezoelectric positioner, and array of vacuum feedthroughs. Each feedthrough must occupy no more than 20 × 20 mm and permit 48 optical fibres and 8 DC connections to pass through the vacuum chamber wall.

Figure 7 shows an example of a vacuum chamber simultaneously satisfying most design constraints, with the exception of loading and 2-D scalability. Cooling lasers enter through the side walls and exit via the roof with the aid of 45 mirrors. An array of roof windows provides unobstructed and close line of sight to all traps. Figure 8 shows a modification of Figure 7 designed to also provide a separate loading region.

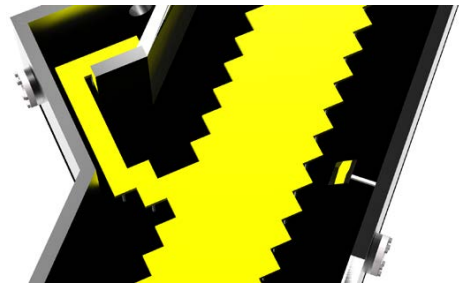

Figure 8: Vacuum chamber divided into a load region (left) and compute region (right) to control contamination. Vertical mirrors strategically attached to the walls and 45° mirrors (rectangles inside the vacuum) ensure that cooling light sheets entering through the chamber windows (rectangles outside the vacuum) can reach the vast majority of the surface of all chips. Chips deepest in the secluded region would be equipped with load slots and ion sources, enabling back-side loading.

---

7  https://www.acktar.com/catagory/MagicBlack

2D scalability could be achieved with a honeycomb vacuum system as shown in Figure 9. The bubbles in the honeycomb corresponding to regions outside the vacuum. These bubbles ensure that the required mechanical strength of the walls does not grow with system size, and that the required alignment precision of the cooling lasers does not grow with system size. For true maintenance and assembly scalability, air locks would need to be built into the system to enable sections of the computer to be removed, replaced, or added, to increase the power of the computer. Each bubble also contains a load region of the form shown in Figure 9, to ensure that all chips are no more than a constant distance from their nearest load region.

The final blueprint design that Universal Quantum has adopted replaces a significant amount of the laser control needed to interact ions with a global microwave field pulse that is applied over all the qubits in the system. Tuning individual qubits in and out of resonance with this microwave field is achieved using a gradient magnetic field that is produced local to each individual qubit by a gradient current carrying wire that is located under zones in each trap that are used for interacting qubits (Weidt et al. 2016).

The system is ultimately designed to realise a large 2D array of qubits that can be interacted on a square grid to produce a collection of surface-code encoded logical qubits. Ion shuttling is utilised to space out the individual ions and allow ions to interact with their neighbours to the north, east, south and west across the entire machine. This is sufficient to achieve a universal error-corrected machine.

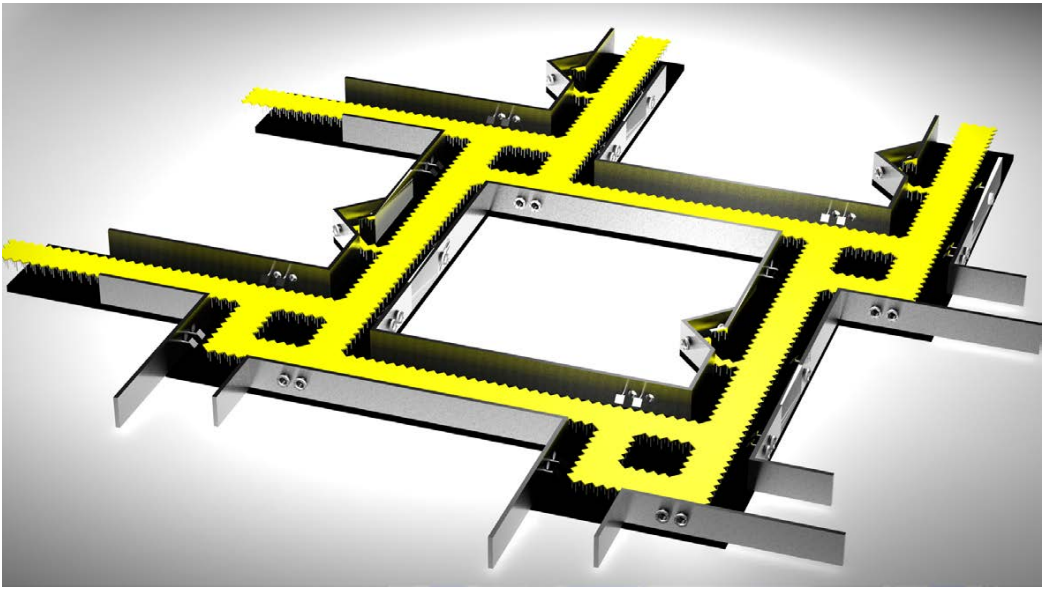

Figure 9: An infinitely extendable honeycomb vacuum system containing an arbitrarily large number of qubits. Note that optical alignment needs to be achieved only over a finite distance. Load regions are included in a regular pattern to ensure scalable ion loading.

The final size of such a honeycomb ion trap system is ultimately dictated by the number of qubits that a quantum algorithm requires. However, for large-scale applications in quantum chemistry, anywhere between 1M and 100M physical qubits will be required, depending on the exact simulation problem. This would imply an extremely large machine, at an effective 5 x 5 mm footprint per physical qubit, you would require a total area of surface traps of 5 × 5 = 25 m$^2$ to 50 × 50 = 2500 m$^2$, and with the necessary honeycomb design the actual physical size of 25–2500 m$^2$ of surface traps would be much larger.

### Nitrogen Vacancy centres in diamond

As with Ion-traps, the ultimate goal for the design is to build a system that can faithfully create the 2D lattice of qubits to produce error-corrected logical qubits with the surface code.[8] In the case of NV diamond, we utilised a slightly different model of the surface code, known as the Raussendorf lattice (Raussendorf, Harrington, and Goyal 2006). This is due to the fact that a diamond-based quantum computing architecture uses a highly probabilistic optical connection to realise quantum gates between individual NV qubits. The Raussendorf lattice is a Measurement Based Quantum Computational (MBQC) version of the surface code. MBQC techniques are particularly useful when an underlying hardware architecture is built using probabilistic gates.

The NV diamond computer is based on qubits which are nitrogen defects within a diamond crystal. The diamond crystal itself provides what is known as a spin vacuum substrate (Aharonovich, Greentree, and Prawer 2011). Essentially, the diamond crystal itself provides the same isolation properties that an actual physical vacuum does for ion-trap technology. The operational temperature of the diamond system is 4K. While still cold, 4K cryogenic technology is far simpler than the dilution refrigeration systems needed for a 30 mK thermal environment for systems such as superconductors. 4K cryogenic technology is so advanced that we are able to effectively launch these sorts of cooling systems into space (Gehrz et al. 2007). In 2003, the Spitzer space telescope was launched by NASA. On board is 360 litre liquid helium cryostat needed to cool instrumentation to approximately 1.5K to look at faint heat signatures from astronomical objects.

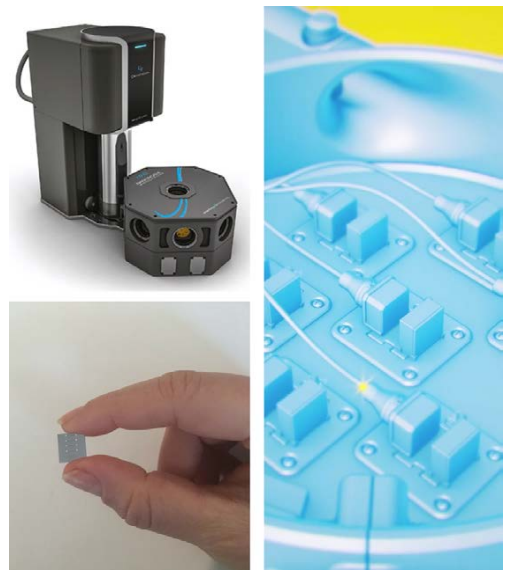

Figure 10: An NV diamond chip-set that is optically coupled, placed inside a 4K helium cryostat system.

---

8  The work in designing an architecture for quantum computation using Nitrogen Vacancy (NV) centres in diamond was performed in collaboration with the Technical University of Vienna and NTT Basic Research Labs (BRL) in Japan (Nemoto et al. 2014).

The diamond-based quantum chip-set is an array of optically coupled nitrogen-defect qubits embedded within a diamond lattice. The chip itself consists of an etched silicon base, with a ultra-thin diamond wafer "glue" on top. The diamond wafer is doped with individual nitrogen atoms separated from each other sufficiently that they don't directly interact. Individual qubits are coupled to each other using a layer of integrated silicon optics that sits above the diamond layer.

Optical pulses are sent between individual nitrogen-defect qubits to enact multi-qubit gates. These optical pulses can, in general, be weak coherent states that are easily produced. The system geometry is spaced out and optimized to allow the control structures for both the NV and optical layer to be fabricated to high accuracy.

Shown in Figure 10 is the device itself. On the right-hand side is a rendering of the microscopic detail of each chip, with multiple qubit arrays (chip-sets) connected to each other with fibre-optic connections. Shown on the bottom left is a single chip, fabricated by the Trupke group at the University of Vienna, containing a micro-cavity system [Figure 11]. On the top left is a commercially available 4K liquid-helium cryostat, similar to the device currently used in the Trupke lab at the University of Vienna.

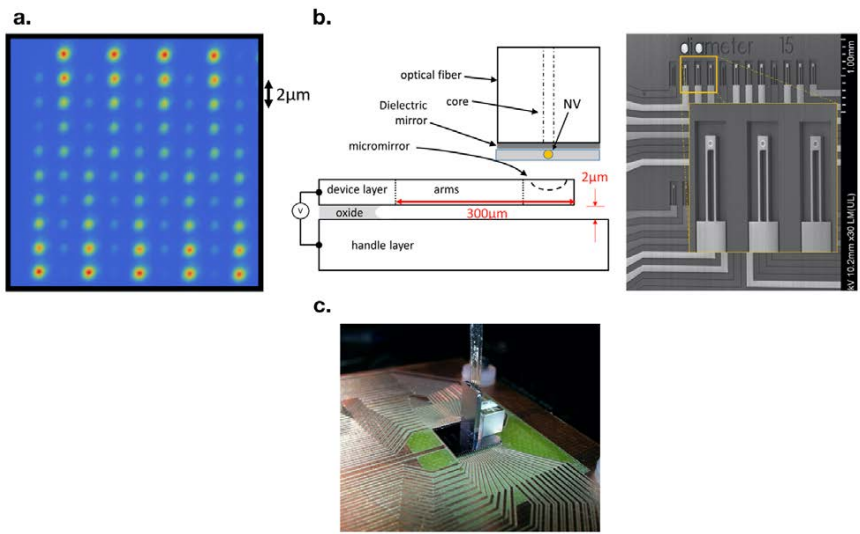

Figure 11: Photos of the three primary components of a diamond quantum chip appropriate for large-scale quantum computing. Fig a: This is a photo of a fabricated sample from Michael Trupke's group at the University of Vienna, implanted with NV-defect qubits in a 9x10 array, with each NV location separated by 2 micrometers. The intensity of each spot indicates the number of NV-centres implanted at each location, where we control the number of NV-defects implanted down each column to a single defect (the dimmest point in each column). This image also demonstrates controlled placement, with each position in our 9x10 array successfully doped. Fig b: Schematic and scanning electron microscope image of an array of micromirrors on cantilevers. The wafer-scale fabrication enables the creation of large numbers of these devices in a single fabrication process. Fig c: Photo of the cavity chip with integrated fibre optics unit connected on top. We have demonstrated controllable and selective coupling between arbitrary pairs of cavities on the chip.

The device consumes approximately 3 to 4 litres of liquid helium over the course of about 12 to 18 months. Losses are mainly though leakage in the closed loop recycling system.

The cryostat system is designed to accommodate multiple quantum chip sets. All the individual physical qubits in the system are connected to each other using integrated silicon optics and fibre connections, and multiple chips can be easily connected to allow quantum information to interact across multiple chip sets.

The NV defect is formed by first growing an ultra-high-purity diamond crystal, where each carbon atom in the lattice is the specific isotope of Carbon-12 and the lattice itself contains no other impurities. Such diamond chips are readily available commercially, with purities above 99.99%. This crystal is then doped, using low-energy ion implantation with single nitrogen atoms. At low enough densities, a single nitrogen atom will form to substitute one of the carbon atoms, and "kick out" an adjacent carbon atom, forming the Nitrogen-Vacancy defect. The diamond lattice itself produces an ultra-clean environment, negating the need for vacuum systems; providing motional stability, and eliminating a large amount of photonic and phononic noise, providing us with very stable, low-decoherence qubits.

In the diamond architecture, the electronic qubit is used as a communications mechanism to allow us to entangle multiple, isolated, qubits with optical photons via etched silicon waveguides and/or fiber optics.

Furthermore, the transition between the ground state of the electronic system and the excited state of the electronic system occurs with a photon at a wavelength of 638 nanometers. This places the transition in the optical frequency range. Unlike many other quantum architectures, the basic physics of the NV-defect provides access using photons at optical frequencies (rather than the more common microwave frequencies). Optical access to an otherwise solid-state qubit system is the key to build a distributed, modular based quantum computer that can scale arbitrarily. By using optical photons and fiber optics, we do not require direct coupling between qubits. This allows us to space out our system, have multiple parts of the computer housed in separate cooling systems and build a machine that can be expanded by simply adding more and more qubit chips as they become available.

*Interacting physically separated NV-qubits*

The mechanism to create entanglement between two physically separated NV-defects makes use of an optical cavity that enhances the interaction between the electronic qubit and an optical field. This is illustrated in Figure 12.

Through a well-known quantum protocol, known as dipole induced transparency, the NV-defect is placed within an optical cavity that resonates at the same frequency as the optical transition of the NV-electronic qubit, when that qubit is only in the zero state. This resonance-matching changes the reflectivity properties of the cavity such that if a photon tries to enter the cavity from outside while the atom is in the zero state, it will be reflected at the entrance to the cavity. If the electronic qubit is in the one state, the photon will enter the cavity and be absorbed or scattered. This quantum mechanical dependence of the reflectivity properties of the cavity gives us a mechanism to produce entanglement between two spatially isolated electronic qubits in two separate NV-cavity systems.

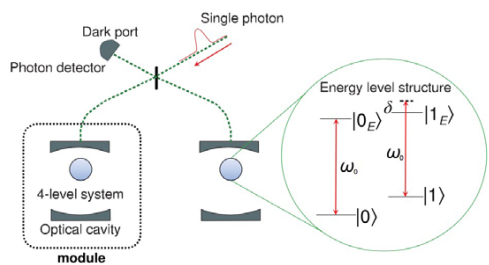

Figure 12: The optical mechanism to entangle two NV-defects. The optical transitions of the respective electronic systems is tuned to resonance with two optical cavities. This resonance condition changes the reflectivity properties of the two cavities dependent on the electronic state. If each electron is placed into a superposition of its two ground states and a single photon that is first split on a beamsplitter is sent to both cavities, there is a finite probability that they will reflect from both cavities and are detected at the photon detector. Because we don't know which cavity the photon reflected from, the resultant electronic state is entangled. The success probability for this scheme is upper bounded at 12.5% when all other components are perfect.

If we take a single photon from an external source (a highly attenuated laser, for example) and first send it through an optical beam splitter, this will place the photon into an equal superposition of heading towards NV-system one or NV-system two. If each of the electronic qubits are placed into an equal superposition of the zero and one states, then the photon will be reflected from each of the cavity systems with a probability of 50%. We then send these reflected photons back through the beamsplitter and with a photon detector, measure if we see a photon come back to us.

87.5% of the time, we do not see any photon with our detector, as the photon may have been absorbed by either one of the two NV-cavity systems or it may not come out the same port of the beam splitter corresponding to the location of the photon detector. However, 12.5% of the time, we will see a "click," indicating that the photon returned to the detector. If we see a "click," then the photon must have reflected off one of the two cavities. However, we cannot say which cavity it actually reflected from. If we cannot ascertain which cavity it reflects off, the final state must be a linear superposition of the two possibilities (photon reflected off cavity one + photon reflected off cavity two). Since the quantum mechanical state of the electron in each of the two NV-cavity systems determines if the photon reflects or not (depending on their respective qubit state), the detection of the photon will result in the electronic states of the two NV-defects being entangled. This is the primary mechanism for NV-defect entanglement.

In the case of the diamond based system, the nitrogen nucleus provides us with a protected memory space to store quantum entanglement while we repeatedly attempt to create new electron–electron entanglement using highly probabilistic quantum gates. This technique is well known in the community and it is commonly referred to as brokered graph state quantum computation.

Using the optical interface, we can attempt to establish an entanglement bond between the electronic qubits in two, physically separate, NV-cavity systems. Conservatively, assuming a connection efficiency of 1%, we require approximately 100 attempts before we are reasonably confident that a connection will be established (>50% chance of a connection after 100 attempts).

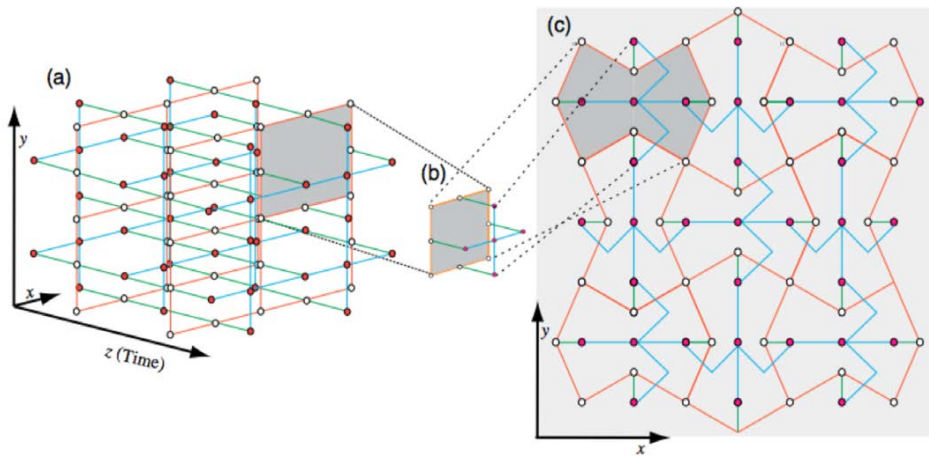

Figure 13: The diamond architecture is based on a 3D cluster state that is a universal resource state for topologically error-corrected quantum computation. Rather than creating a full 3D lattice, we create a 2D+1 lattice on a planar lattice of NV-qubits. This allows for arbitrary quantum algorithms that are protected from errors using powerful topological codes.

Given the intrinsic speed of this system, these 100 attempts can be completed in approximately 3 microseconds (Nemoto et al. 2014). If we were to try to connect a third electronic qubit after this initial successful connection, any failed attempt would destroy the entanglement we had previously established. Hence, before we attempt to connect a third electronic qubit into our entangled state, we use the hyperfine interaction between the electrons and their respective nitrogen nuclear qubits to transfer the entanglement.

If a successful entanglement bond is transferred to the nuclear qubits, this "frees" the electron qubits to be used again to probabilistically create another entanglement bond with other electronic qubits in other NV-cavity systems without the possibility of destroying the entanglement we have already created. The nuclear system acts as a protected memory (or broker) to allow for further creation of electron/electron entanglement. The protected nuclear memory allows for extremely inefficient optical connections without adversely impacting the error rates associated with the creation of NV/NV entanglement or slowing the computer down by such a degree as to make large-scale computation impractical.

*Creating an entangled resource state for error-corrected computation*

The basic mechanism for NV/NV entanglement can be used to create a cluster state of arbitrary size. Hence the diamond architecture can be used to build a universal resource state for large-scale, error-corrected quantum computation.

Embedding topological error-correction codes into our computational model simply requires us to produce an entangled resource state that can be used to mimic surface code, error-corrected logical qubits. This requires the creation of a 2D+1 dimensional cluster state, known as the Raussendorf lattice (Raussendorf, Harrington, and Goyal 2006). The structure of this state is illustrated in Figure 13.

The cluster state we require is abstractly a 3D state. The cross section of this state represents the size of the computer — how much space we have in terms of error-corrected qubits — while the third dimension represents time steps available in our computation — the longer the third dimension is, the more steps we have available. Creating this 3D cluster state directly would introduce a temporal scaling to our architecture — we would need more physical qubits to enact more error-corrected gates — but we do not need to create the entire third dimension in physical space. In fact we only need to create two layers at a time along the temporal axis of the cluster state. Hence the cluster we need to prepare is the 2D cross section, plus one additional layer along the temporal axis (i.e. a 2D+1 cluster state).

Figure 13c illustrates the actual physical layout of the cluster state that we need to produce, assuming each of our physical NV-defects lies on a 2D plane. We basically take two sequential cross-sections of the 3D cluster and "pancake" it down to 2D. This results in a 2D qubit layout that requires next-to-nearest-neighbour connections (the colour coding in Figure 13 is not relevant to this discussion), but the optical connectivity of this architecture allows for these longer-range connections.

Creating a cluster state with this general structure provides an appropriate universal resource state for fault-tolerant, topologically error-corrected, universal computation. The size of the computer and the strength of error correction is only related to the actual 2D size of the array shown in Figure 13c. The array in Figure 13c consists of cells in the 3D cluster state, and a large-scale quantum algorithm (for example factoring) would require arrays of the order 10,000 × 10,000 cells (Nemoto et al. 2014).

Because a diamond system is optically connected and cooling infrastructure is comparatively simple when compared to large vacuum systems or dilution refrigeration, a highly distributed diamond system is expected to be simpler to scale. While such a machine would still be large, it would not need to be a carefully interconnected high-vacuum systems or ultra-large dilution refrigerators.

### Superconductors

The motivation behind the Japanese superconducting micro-architecture is a well-known problem within superconducting quantum computing known as "the wiring problem." This micro-architecture takes a fundamentally different structural approach to eliminate this problem without changing any other issue related to large-scale operation of a superconducting quantum computer.[9]

Superconducting qubit systems have arguably emerged as the leading platform for large-scale computing architectures. Not only have we seen significant advances in recent years in reliable fabrication and control technology, but the quality of the qubits themselves has increased by many orders of magnitude. Superconducting qubit systems have demonstrated physical error rates close to (and in some cases, below) the fault-tolerant threshold for surface-code-based error-correction techniques (Barends et al. 2014), and multi-qubit arrays have been fab-ricated and tested by many groups worldwide. Private investment in superconducting quantum computing technology has also exploded, with companies such as Google, IBM, Alibaba, Intel, Rigetti and others now actively and aggressively promoting and supporting the platform. The first demonstration of quantum supremacy was demonstrated with Google's 53-qubit, Sycamore superconducting processor, and IBM have deployed over 20 superconducting quantum computers through the cloud on the IBMQ network.[10]

---

9  The third design discussed is a modified design for a superconducting quantum computer that was designed in collaboration with the group of Jaw Shen-Tsai at the Japanese national laboratories, Riken and Tokyo University of Science (Mukai et al. 2020; Kwon et al. 2020).

Scaling these systems to the level needed to achieve error-corrected, commercial applications will require integrated chipsets containing of order 1000 physical qubits or more in a surface code error-corrected logical qubit and this presents certain technological challenges. One of the most significant is the so-called wiring problem. This is the fact that a large, error-corrected array of superconducting qubits requires a 2D qubit chipset that allows for nearest-neighbour couplings and that qubits within the centre of such a chip cannot be directly accessed for the fabrication of bias lines, control lines and readout machinery.

The current consensus within the superconducting community is that the control wiring for such chips should be fabricated in the third dimension, utilizing several techniques to place bias, readout and control wires orthogonal to the plane of the chipset itself. This technique has shown promise (Rosenberg et al. 2017; Foxen et al. 2017), but it is very unclear if these control fabrication techniques are compatible with maintaining high-fidelity operations. The largest concern is the ability to reduce cross-talk and control line contamination of neighbouring qubits to the degree necessary to achieve fidelities of 99% or higher across the chip.

This new micro-architecture was designed specifically to side-step this issue. We demonstrated that a pseudo-2D arrangement of superconducting qubits — completely compatible with surface code based error-correction — can be constructed in a physical bi-linear arrangement of superconducting qubits. This bi-linear array allows for each physical qubit to be biased, measured and controlled using wiring that remains in-plane with the chipset — eliminating completely the need for 3D control line fabrication.

To achieve this new architecture we introduced small air-bridges within the resonators coupling together individual qubits. These air-bridges allow us to create a criss-cross resonator design that allows us to create the pseudo-2D qubit arrangement. We demonstrated that the resonator quality and crosstalk is not adversely affected by the introduction of these air-bridges and that we can anticipate no adverse effects on the architecture by moving to this new design.

*A standard superconducting quantum microarchitecture*

To maintain compatibility with quantum error-correction codes, a minimal design for a superconducting chipset is a 2D nearest-neighbour interacting array. Google's Sycamore processor is a 54-qubit planar wafer design. Illustrated in Figure 6a is a schematic of the qubit (gray crosses) layout, where 2D nearest-neighbour interactions are mediated by adjustable couplers (blue boxes). The four adjustable couplers allow switchable quantum gates to be implemented between a given qubit and its four nearest neighbours. In Sycamore, a 54-qubit chipset was initially designed, but one of the 54 qubits was non-functional after fabrication (white cross).

---

10  https://www.ibm.com/quantum-computing/systems/

Figure 14b shows an actual photograph of the 10mm Sycamore chip. The central square region is the actual chipset, while emanating from the central processing region are control lines that are used to bias, control and measure each of the 54 qubits and control each of the 88 adjustable couplers (both built from transmons[11]).

The packaging of Sycamore (i.e. the control lines that need to be fabricated from each transmon to the boundary of the chipset) is already quite dense and complex. In the context of the surface code error-corrected qubit, an array of 54 qubits would only be sufficient to construct a distance $d = (\sqrt{N} + 1)/2 = 4$ logical qubit (using a standard square planar code configuration (Horsman et al. 2012)). This is a very small amount of error correction and would not be sufficient for any large-scale quantum algorithm.

If we expanded to a $N = 1521$ or $N = 2401$ chipset, corresponding to distances $d = 20$ and $d = 25$ code respectively, it would be simply too dense to have enough physical space on the chipset for bias, control and measurement lines for both the transmons acting as qubits and the transmons acting as adjustable couplers. (A $N = 1521$ qubit chipset would require over 3000 additional transmons as couplers, for a total of over 4600 transmons per chip.)

Planar wiring is therefore not a viable method to scale a microarchitecture of this type. Instead, the common method is to envisage control lines to be fabricated perpendicular to the chip plane. These three-dimensional wiring technologies generally consist of techniques such as flip-chip bonding, pogo pins, and through-silicon vias (TSVs) (Barends et al. 2014; Takita et al. 2017; Reagor et al. 2018; Chou et al. 2018; Bejanin et al. 2016; Vahidpour et al. 2017; Foxen et al. 2018; Rosenberg et al. 2017).

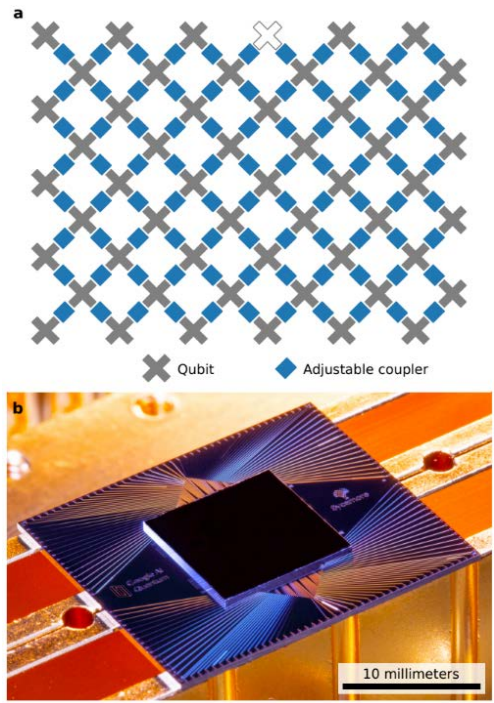

Figure 14: The Google Sycamore processor (Arute et al. 2019). Fig a represents the layout of 54 qubits (grey crosses, one white cross for a non-functioning qubit) in a 2D nearest neighbour microarchitecture. Each blue box represents a transmon that is used as an adjustable coupler. Fig b is a photo of the Sycamore chip. You can see control wires that emanate from the central square region containing the qubits.

While it is still unclear if complex 3D wiring will affect the performance of the underlying chipset — fidelities of superconducting chipsets are very much on the boundary of the threshold for what is required of error-corrected system — it is expected to be a source of fabrication and cross-talk noise that would be good if it could be eliminated completely.

---

11  A transmon is a type of superconducting charge qubit designed to have reduced sensitivity to charge noise. [Ed.]

*The superconducting air-bridge resonator*

The key to this new micro-architecture design was introducing an air-bridged superconducting resonator. An air-bridged resonator is a standard Coplanar Waveguide Resonator (CWR) that contains a break in the waveguide and a literal bridge that connects the two halves of the waveguide. Figure 15 shows a scanning electron microscope (SEM) image of a superconducting air-bridge that was fabricated by the Google group (Chen et al. 2014).

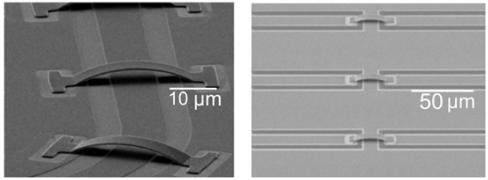

Figure 15: A scanning electron microscope (SEM) image of a series of air-bridges fabricated by the Google quantum team (Chen et al. 2014).

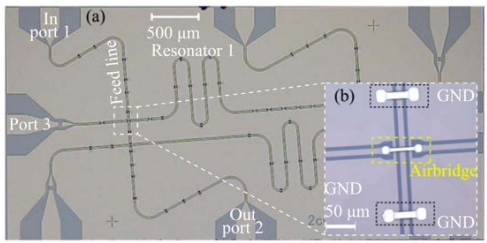

Figure 16: A SEM image of four crossed resonators, fabricated by the Tokyo University of Science group as part of this microarchitecture redesign. From (Mukai et al. 2020).

The existence of a bridge allows us to consider a cross resonator design. This is where a CWR is fabricated under a bridge that connects a second CWR fabricated in an orthogonal direction. Illustrated in Figure 16 is a SEM image of the actual device that the Tokyo University of Science team fabricated as part of this project.

This prototype device consisted of four resonators: two run vertically across the chip and two run horizontally. The resonators have four crossing points, where an air-bridge is used so that one resonator travels under the other. Notice that air-bridges are used in multiple locations around the CWR chip, to ensure that there is a common electrical ground. If certain conducting islands were isolated on the chipset, it would create a potential difference between regions and consequently create stray capacitances.

Along the horizontal resonators, a series of air-bridges were fabricated along its length. These air-bridges did not have a secondary resonator passing underneath and were fabricated to test whether there was any level of resonator degradation as a function of the total number of air-bridges along a resonator.

Figure 17 illustrates experimental data of the prototype air-bridged system. In Figure 17a we plot the infidelity of a resonator-mediated quantum gate between two superconducting qubits as a function of the quality factor of the resonator. The horizontal dotted line represents the error rate (infidelity) reaching the level where error correction is viable (approximately 0.7%) and the vertical line is the experimental measured quality factors of the test resonator system with between 15 and 20 air-bridges. The blue curve lies under the horizontal line when it intersects with the vertical line, mean-ing that a resonator containing between 15 and 20 air-bridges can still be used to enact quantum gates between two superconducting qubits at an error rate below the surface code threshold. (top) 25 mm (bottom), excluding headers and footers.

resonators for a pair of two-qubit interactions — each interaction using one of the two resonators — still needs to be performed when the full four-qubit system is fabricated.

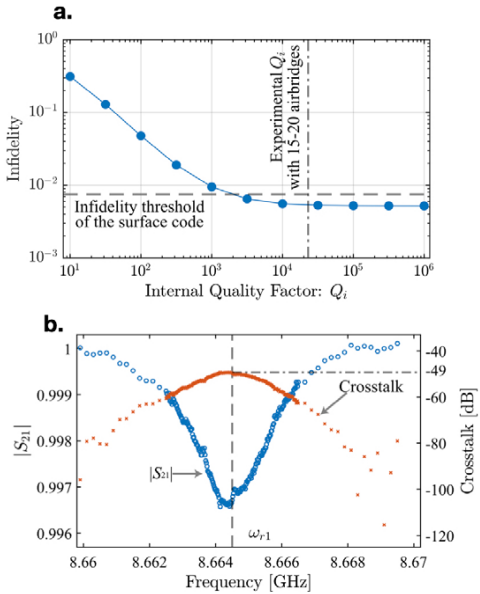

Figure 17: Fig a. illustrates the infidelity of a resonator mediated quantum gate between two superconducting qubits as a function of the quality factor of the resonator. Fig b. illustrates the resonance of the air-bridged CWR (blue) and the measured noise that contaminates the orthogonal resonator (red). From (Mukai et al. 2020).

Figure 17b is a plot of the resonance dip of the CWR (blue) and the measured crosstalk to the orthogonal resonator that passes under the air-bridge (red). At resonance, the maximum measured crosstalk was approximately −49dB. This demonstrates that resonator crosstalk for a crossed resonator system is effectively non-existent. Potential induced phase-shifts on the resonator mode during simultaneous use of the two

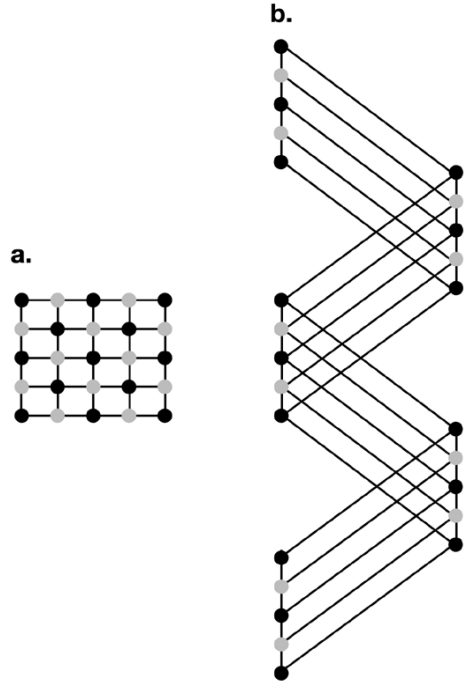

Figure 18: A new bi-linear microarchitecture formed from the standard 2D layout of qubits needed for surface code error correction. Fig a illustrates a standard logical qubit that requires a grid. Fig b illustrates the new bi-linear arrangement where couplings between columns are achieved using air-bridged crossed CWRs.

*A bi-linear microarchitecture for superconducting quantum chips*

Once air-bridges can be introduced on resonators, enabling the ability to cross resonators, we can redesign the standard 2D qubit layout with nearest-neighbour interactions to a bi-linear array of super-conducting qubits that interact through a series of crossed resonators.

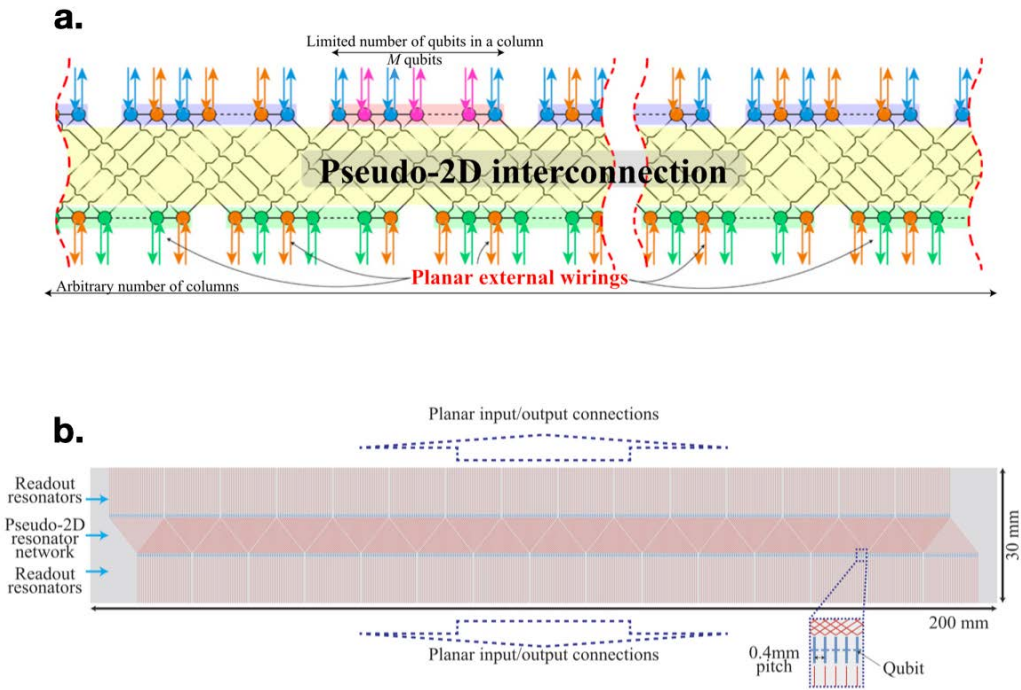

Figure 19: Physical layout of the new microarchitecture. Fig a: An arbitrarily long but fixed-width surface code can be created using a bi-linear arrangement of superconducting qubits. The fixed width of the surface code ensures that the air-bridged resonators have a finite length and number of air-bridged crossings. Each superconducting qubit can be accessed in the plane for the control, initialisation, and readout technology. Fig b: A design of a logical qubit chip consists of 30×30 physical qubits, encoding a $d = 15$ logical qubit. Qubits are depicted as blue crosses (inset) and resonators are as red lines. All the external input and output connections can be achieved by the conventional planer wiring technology. The resulted chip size is approximately 30 mm × 200 mm rectangular. From (Kwon et al. 2020).

The modification to the microarchitecture is shown in Figure 18. Figure 18a shows a standard 5 × 5 surface code array that corresponds to a distance $d = 3$ error-correction code. As you can see, the qubits within the interior become difficult to access for fabrication of bias lines, control and readout lines.

If we take each column of this 2D array and lay them out in two distinct columns, alternating columns from the original 2D patch, we can create a bi-linear array, as shown in Figure 18b. Required qubit interactions within a column do not change, but interactions between columns now become longer range and they cross. These interactions can be achieved using air-bridged resonators.

The significant difference between these two layouts is that, in the bi-linear array, there is now lateral planar access to every qubit within the computer. Qubits that used to be buried within the centre of the 2D lattice are now placed along one of the columns in the bi-linear array, where bias, control and measurement lines can be fab-ricated and placed immediately alongside. This solves the 3D wiring problem by introducing a new component, the air-bridged resonator, which was demonstrated not to adversely affect the performance of a quantum gate between two superconducting qubits.

As each column is alternatively placed on each side of the bi-linear array, the number of air-bridges is dictated by the number of qubits within a column. As shown in Figure 19a, each block (both green and blue) corresponds to a single column of $M$ qubits. The number of crossing points for air-bridges is $M - 1$ and as we can alternate on which resonator the air-bridge is actually fabricated, the total number of air-bridges on a single resonator is $(M - 1)/2$. Before discussing the total length of the superconducting chip-set, we need to detail how qubits are arranged

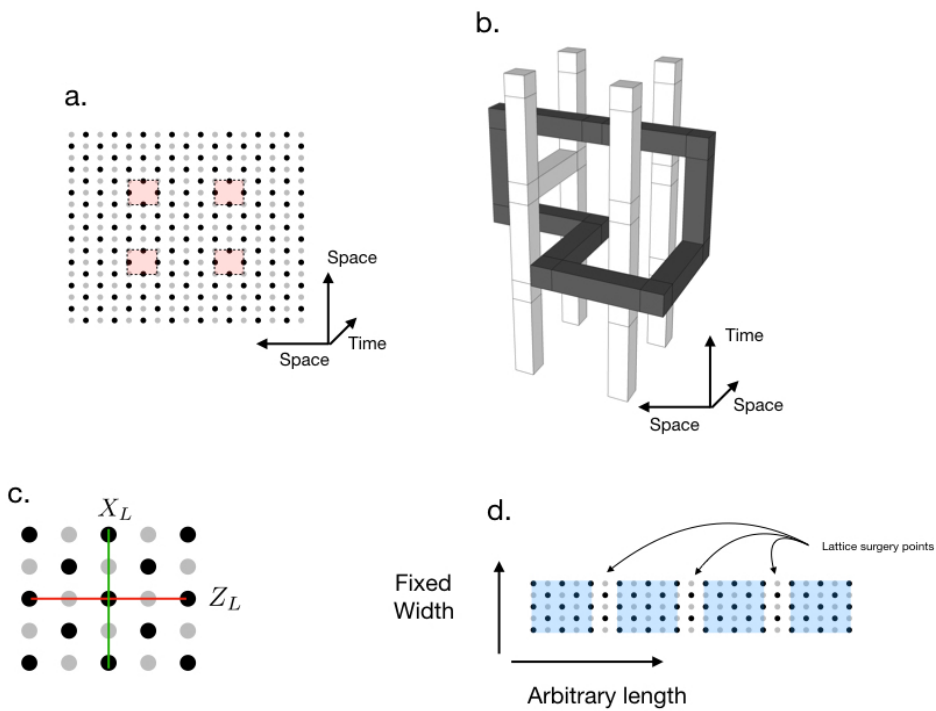

Figure 20: Braid and lattice surgery based logic on a standard 2D surface code. Fig a is how a set of two logical qubits, encoded with a distance d = 3 surface code are introduced as four pairs of defects, where a defect is defined as a region of the surface code where physical circuits are switched off (shaded in red). Fig b is a logical CNOT operation enacted over two logical qubits (the two pairs of white defects). Time is represented vertically and, as the circuit is executed, defects are moved to complete topological braids. A quantum circuit is consequently represented as a geometric figure, with a cross-section related to the number of physical qubits needed and the third axis representing the time needed to complete the logic gate. Fig c is a single planar code logical qubit at d = 3. In this situation, logical qubits are isolated from each other until logic operations via lattice surgery are enacted. Fig d illustrates the associated logical qubit layout for lattice surgery logic. Each logical qubit is illustrated as the blue shaded region, with an extra column of physical qubits used as the merge/splitting points for lattice surgery logic. Significant physical resources are saved using lattice surgery compared to defect-based encoding.

in this system so that logic operations can be performed in the error-corrected system.

In the surface code, there are two predominant models for performing error-corrected logic. The first techniques is known as topological braiding, and was the original formalism for error-corrected logic (Raussendorf, Harrington, and Goyal 2006; Fowler and Devitt 2012; Fowler, Devitt, and Jones 2013; Paler, Devitt, and Fowler 2016; Paler et al. 2012; Devitt et al. 2013).

The layout of qubits for braided logic is a full 2D array of qubits as shown in Figure 20a. In this array, we have four regions (coloured in red) where the error-correction procedures that usually occur across the lattice are simply not performed ("switched-off"). This creates holes or "defects" within the code. These defects are effectively degrees of freedom in the code-space that can be used to encode multiple qubits of logical information.

The error-correction strength of the information encoded within these defects is determined by the circumference of the defect and the separation in the lattice between defects. In Figure 20a the number of data qubits (black qubits) that circumscribe a defect region is four and the minimum number of data qubits between any two defect regions is three. Consequently the code distance used to encode the information is $d = 3$ (the minimum of the two). For various technical reasons (Raussendorf, Harrington, and Goyal 2006), a logical qubit of information is represented by two defects. Hence Figure 20a represents a computer encoding two logical qubits with a distance $d = 3$ code.

Logic operations are performed by changing the locations of these defect regions over time and tracing out a space/time geometry of world lines for these defect regions. An example of this is given in Figure 20b. The XY-plane of this diagram represents the spatial 2D lattice of Figure 20a, where the four white pillars represent four of the defects. Time is represented by the Z-direction. As the computer evolves over time, the defect regions in the lattice are moved (by switching off and on parts of the 2D lattice) to trace out the geometric structure of Figure 20b. This enacts a gate operation. In this case, a two-qubit logically encoded CNOT operation. The control qubits are represented by the two white pillars to the left, the target qubit is the two white pillars on the right. The input for the circuit is at the bottom of the image while the output is at the top. The details of how these gates actually enact logic operations can be found in several references (Fowler and Devitt 2012; Fowler, Devitt, and Jones 2013).

The main point from Figure 20a is that the computer needs to be effectively a large square 2D array of physical qubits, with defects placed throughout the array. This is not compatible with the new bi-linear array microarchitecture where we are limited to the total length of a column in the 2D lattice which dictates the number of air-bridged resonators needed by the design.

This can be solved by moving to a different methodology of fault-tolerant logic called lattice surgery (Horsman et al. 2012). In the lattice surgery model, logical qubits are simple square patches of surface code, determined by the distance of the underlying quantum code. A square $M \times M$ patch corresponds to a code distance of $d = (M + 1)/2$ [Figure 20c]. Each square patch is laid next to each other This effectively means that our arrangement of logical qubits in the computer becomes a Linear

Nearest Neighbour array (LNN) [Figure 20d]. We have a sufficient number of rows in the design to encode a single logical qubit and we have as many columns as necessary to house the total number of logical qubits in the computer.

In lattice surgery, isolated square patches of planar code are interacted along a boundary to enact multi-qubit logic gates. This reduces the overall physical resource cost of each logical qubit and several results now suggest that lattice surgery techniques will always be more resource efficient when implementing large-scale algorithms (Herr, Nori, and Devitt 2017; Litinski and Oppen 2018; Fowler and Gidney 2018).

For a single logical qubit encoded with the planar code, a square 2D array of physical qubits is needed. For a distance $d$ quantum code, a $(2d - 1) \times (2d - 1)$ array of physical qubits is sufficient. Illustrated in Figure 20 is a distance planar code requiring 25 physical qubits with the associated logical bit-flip ($X_L$) and phase-flip ($Z_L$) operators illustrated. In Figure. 20d, we illustrate the same LNN logical layout of planar code logical qubits that require less physical resources than defect-based logical qubits. In Figure 20d there is an additional column of physical qubits that are spacers between each encoded qubit that is required to perform the lattice surgery operations. It should be noted that the current methods for circuit compilation using lattice surgery still assume a 2D nearest-neighbour arrangement of logically encoded qubits (Herr, Nori, and Devitt 2017; Fowler and Gidney 2018). Compilation into this pseudo-LNN logical structure will require modifications over current techniques (Herr, Nori, and Devitt 2017). However, this won't adversely impact the physical structure of this new architecture.

For a very large error-correcting code, $d$ can be of the order of 15–20, requiring an array containing 29–39 rows of qubits with 29–39 columns, per logically encoded qubit. Consequently, for a quantum computer containing $N$ logical qubits at distance 15 on the planar code, we would utilize an array of $29 \times (29N + (N - 1))$. Here, 29 is the number of qubits in a column, and $29N$ is the number of columns in the array for each logical qubit, and the extra factor of $(N - 1)$ is the spacing region between each logical qubit needed for lattice surgery. This would translate into a bi-linear array, as shown in Figure 19b of $N(2d-1)(2d-2) = 29 \times 30N$, with each set of air-bridged cross-resonators having at most $[(2d - 1)/2] = 15$ crossings. The factor of ½ comes about due to the fact that alternate resonators can be chosen to contain an air-bridge. Hence, while 29 crossings are at required at most, a given resonator will only contain half that number of air-bridges.

In Figure 19b we illustrate a much larger array, where each individual qubit has wiring access from either above or below the bi-linear array. The cross-resonator network containing the air-bridges are contained within the centre, and allows for the bi-linear array to operate as if it were a long, rectangular 2D array of physical qubits. A system of this size would represent a single superconducting chipset, containing $30 \times 30 = 900$ physical qubits, encoded into a single error-corrected qubit. This would correspond to a distance of $d = 15$ quantum code, sufficient to correct for up to seven physical errors in each error-correction cycle. Estimates on the physical size of this chipset will be 200 mm × 30 mm.

This new micro-architecture for superconducting quantum computers effectively eliminates the problem of wiring up a mas-sive 2D array of physical qubits. Instead, we translate the qubit chip into a bi-linear array of physical qubits that have direct lateral access for bias, control and measurement wiring. While superconducting chipsets have not yet reached the scale where a shift to this new micro-architecture is needed, as these systems scale further a new approach to the underlying structure will need to be adopted by essentially all manufacturers of superconducting quantum computers.

### Designing systems of the future

As nascent quantum chipsets further progress, redesigns and additions to these blueprints will undoubtably occur. The basic building blocks for each of these quantum computing platforms have been demonstrated and there is a clear conceptual pathway to a large-scale system, capable of universal, error-corrected quantum computation.

It is anticipated that many of the challenges that lie ahead are related to the ability to further decrease error rates through improvements in fabrication and control, and solve the problems related to how we scale systems under the constraints of environmental infrastructure such as dilution refrigeration systems and vacuums.

There have been significant advancements in the construction of error-correction protocols and quantum resource optimisation. The surface code still remains the preferred technique for error-correction in experimentally realisable large-scale systems, and physical qubit resources continue to drop as theorists develop new and improved methods for error-corrected logic operations and algorithmic compilation.

The timeframe of when a fully error-corrected system will become available to implement scientifically or commercially useful quantum computing systems is still unknown, but for many platforms the initial ingredients have been demonstrated and it is becoming clear that engineering challenges and capital may be much more significant than any fundamental issues of quantum physics.

### Acknowledgements

The work reviewed and referenced in this paper could not have been achieved without fantastic collaborations with theorists and experimentalists, physicists and computer scientists. I would like to acknowledge the ion-trap group at the University of Sussex, the NV-diamond groups at the University of Vienna and the Technical University of Vienna and the Superconducting quantum research teams at Toyko University of Science and Riken, Japan, for their fantastic collaborations on the blueprints reviewed. Much of the initial work on architectures and system design was performed in collaboration with K. Nemoto and W.J. Munro of the National Institute of Informatics and NTT labs, Japan, and I would like to acknowledge my many collaborators who worked with me on error-correction, compilation and optimisation for large-scale machines such as A.G. Fowler, A. Paler, D. Herr, D. Horsman, R. Van Meter, A.D Greentree, T. Tilma and A.M Stephens.

While I have attempted to cite many other results from others on architecture development in all of the systems examined in this paper, I am sure I have missed many wonderful papers. The context of this manuscript did necessitate it being "me"-centric and that in no way should be interpreted to downplay the extraordinary work that exists within the field of quantum architectures and system design.

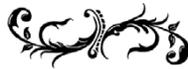